\documentclass[9pt,twocolumn,twoside]{pnas-new}
\templatetype{pnasinvited} 
\usepackage[export]{adjustbox}
\setboolean{displaywatermark}{false}
\begin{document}
\title{The Neuron as a Direct Data-Driven Controller}

\author[a]{Jason Moore}
\author[b]{Alexander Genkin} 
\author[b]{Magnus Tournoy}
\author[b]{Joshua Pughe-Sanford}
\author[c]{Rob R. de Ruyter van Steveninck}
\author[a,b,1]{\\Dmitri B.\ Chklovskii}
\affil[a]{Neuroscience Institute, NYU Langone Medical School}
\affil[b]{Center for Computational Neuroscience, Flatiron Institute}
\affil[c]{Physics Department, Indiana University}

\leadauthor{Moore}
\authorcontributions{DC designed research; DC, JM, AG, MT, JPS, 
performed research, JM, AG analyzed data, RRRS contributed data; DC, JM wrote the paper}
\authordeclaration{Authors declare no competing interests.}
\equalauthors{}
\correspondingauthor{\textsuperscript{1}To whom correspondence should be addressed. E-mail: chklovskii@gmail.com}

\keywords{Neuron $|$ Control $|$ Dynamics $|$ ...}

\begin{abstract}
 In the quest to model neuronal function amidst gaps in physiological data, a promising strategy is to develop a normative theory that interprets neuronal physiology as optimizing a computational objective. This study extends the current normative models, which primarily optimize prediction, by conceptualizing neurons as optimal feedback controllers. We posit that neurons, especially those beyond early sensory areas, act as controllers, steering their environment towards a specific desired state through their output. This environment comprises both synaptically interlinked neurons and external motor sensory feedback loops, enabling neurons to evaluate the effectiveness of their control via synaptic feedback. Utilizing the novel Direct Data-Driven Control (DD-DC) framework, we model neurons as biologically feasible controllers which implicitly identify loop dynamics, infer latent states and optimize control. Our DD-DC neuron model explains various neurophysiological phenomena: the shift from potentiation to depression in Spike-Timing-Dependent Plasticity (STDP) with its asymmetry, the duration and adaptive nature of feedforward and feedback neuronal filters, the imprecision in spike generation under constant stimulation, and the characteristic operational variability and noise in the brain. Our model presents a significant departure from the traditional, feedforward, instant-response McCulloch-Pitts-Rosenblatt neuron, offering a novel and biologically-informed fundamental unit for constructing neural networks.
\end{abstract}

\dates{This manuscript was compiled on \today}
\doi{\url{www.pnas.org/cgi/doi/10.1073/pnas.XXXXXXXXXX}}

\maketitle
\thispagestyle{firststyle}
\ifthenelse{\boolean{shortarticle}}{\ifthenelse{\boolean{singlecolumn}}{\abscontentformatted}{\abscontent}}{}

\significancestatement{
We build upon and expand the efficient coding and predictive information theories, presenting a novel perspective that neurons not only predict but also actively influence their future inputs through their outputs. We introduce the concept of neurons as feedback controllers of their environments, a role traditionally considered computationally demanding, particularly when the dynamical system characterizing the environment is unknown. By harnessing an advanced data-driven control framework, we illustrate the feasibility of biological neurons functioning as effective feedback controllers. This innovative approach enables us to coherently explain various experimental findings that previously seemed unrelated. Our research has profound implications, potentially revolutionizing the modeling of neuronal circuits and paving the way for the creation of more sophisticated, biologically inspired artificial intelligence systems.}

Despite the wealth of mechanistic insights into neuronal physiology, constructing generalizable models of brain function remains a formidable challenge in neuroscience. This difficulty largely stems from the inherent variability of biological neurons, characterized by an array of challenging-to-quantify parameters like ion channel densities. A promising strategy to overcome this challenge involves developing a normative theory of neuronal function, conceptualizing neuronal physiology as an optimization of a computational objective. Such a normative theory can potentially mitigate the limitations posed by scarce physiological data through a focus on the functional integrity of computational models.

Shining examples of such a normative approach are the efficient coding and predictive information theories. Efficient coding \citep{barlow2001redundancy,srinivasan1982predictive,atick1992could,van1992theoretical,simoncelli2001natural,rao1999predictive,jun2022efficient}, by maximizing transmitted information under physical constraints, views spike-triggered averages (STAs) as optimal feedforward filters and rationalizes their adaptation with input statistics. Predictive information theories \citep{palmer2015predictive,rust2021remembering,wang2021maximally,tishby2000information,chalk2018toward}, by optimizing the encoding of future-relevant information, have demonstrated quantitative congruence with experimental observations in early sensory areas. These theories apply beyond these areas, as evidenced by the adaptive nature of feedforward filters in other neuronal types \citep{bryant1976spike,mainen1995reliability}.

However, this perspective does not fully account for certain physiological attributes of neurons. Our analysis reveals that neurons adapt not only their feedforward filters but also their spike-history-dependent (feedback) filters, suggesting a functional role beyond basic housekeeping operations like sodium channel deinactivation during refractory periods. Furthermore, whereas current injections in neurons with identical high-variance waveforms produce consistent spike trains, constant current injections result in more variable outputs \citep{mainen1995reliability}. Neither feedback filter adaptation, nor inconsistent response to constant current injections are predicted by efficient coding.

While prediction remains a crucial aspect of neuronal computation beyond early sensory areas, it likely isn’t the sole computational objective. Neurons, particularly in motor and pre-motor areas, are tasked with not only forecasting but also influencing future states of the external environment through precise control signals. Additionally, the pervasive presence of feedback loops in the brain \citep{bell1999levels,song2005highly,varshney2011structural,winding2023connectome} underscores that neuronal outputs often modulate their own inputs physiologically. 

\begin{figure}[hbtp]
     \centering
      \includegraphics[width=0.49\textwidth]{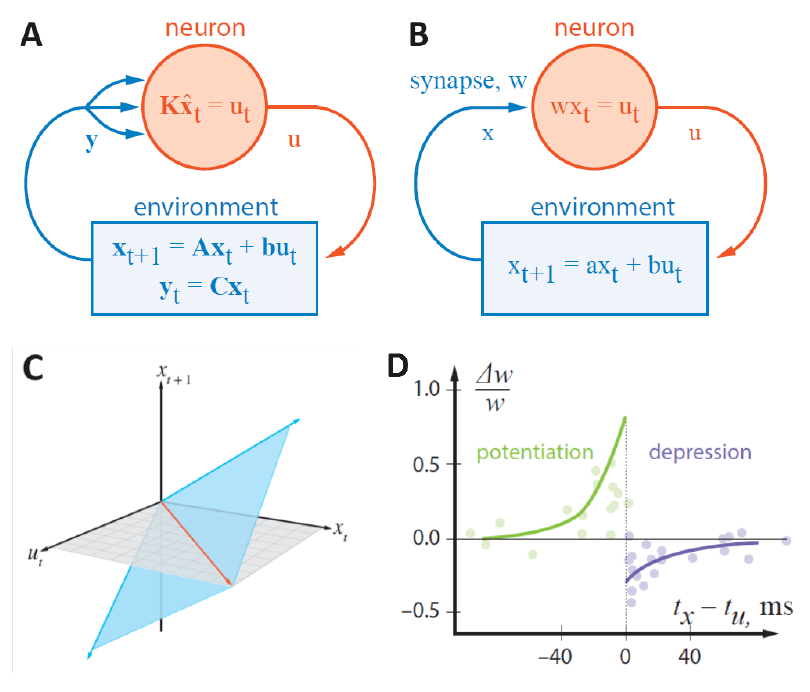}
        \caption[width=0.49\textwidth]{{\bf A}: A schematic representation of a neuron as a feedback controller in a closed loop. {\bf B}: A scalar fully observed dynamical system controlled by tuning the weight of a synapse, $w$, in the control law. {\bf C}: The subspace of valid pairings of observations and controls (blue plane) is spanned by the previously observed states (blue vectors). The intersection of the valid dynamical subspace with the $x_{t+1}=0$ plane defines the control law (red line). {\bf D}: Spike-timing dependent plasticity (STDP): the relative change in the synaptic weight, $\Delta w/w$, vs. the time interval between the pre- and post-synaptic spikes, $t_x-t_u$, showing the potentiation (causal) and depression (anti-causal) windows \citep{markram1997regulation,bi1998synaptic,zhang1998critical}.}
    \label{fig1}
\end{figure}

These observations have led us to expand the predictive neuron model, incorporating optimal feedback control into the normative framework. We posit that neurons, especially those beyond early sensory areas, act as feedback controllers, aiming to steer their environment toward a desired state, as depicted in Fig. \ref{fig1}{\it A}. The neuronal environment encompasses both the circuits of interconnected neurons and external motor sensory loops, allowing the neuron to assess control efficacy through synaptic feedback.

At first glance, the task of being a feedback controller may seem daunting for a neuron. To begin with, the dynamics of its environment are not known to the neuron a priori, necessitating learning them from data. Traditional system identification methods tackle this by deducing dynamic parameters (e.g., parameters {\bf A}, {\bf b}, and {\bf C} in linear state-space models, as illustrated in Fig. \ref{fig1}{\it A}) from historical observations and control signals \citep{katayama2005subspace}. These parameters form the basis for deriving a control law that optimizes specific objectives, like optimal or robust control \citep{aastrom2021feedback}. When dealing with low-dimensional or noisy observations, the control law needs to be based not on immediate observations, {\bf y}, only but on an estimated state, $ \hat{\bf x} $, \citep{aastrom2021feedback} derived from past data—a process known as output feedback control and Kalman filtering (Fig. \ref{fig1}{\it A}). Although, for linear dynamics, the above tasks have known solutions \citep{katayama2005subspace,aastrom2021feedback}, they are computationally too demanding for a single neuron to perform or even to represent the dynamic parameters explicitly.

To implement a biologically plausible feedback controller, we adopt the novel DD-DC framework \citep{willems2005note,de2019formulas,markovsky2022data}. The crux of DD-DC is to sidestep the explicit representation of the controlled dynamical system and the explicit inference of the latent state, instead directly mapping observations to control signals. This mapping is learned from historical pairings of observations and control signals. In scenarios where the remainder of the loop is represented by a linear dynamical system of order $n\geq 1$, with scalar input (control signal) and output (observations), this relationship is characterized by an Auto-Regressive Moving Average (ARMA) process, as depicted in Fig. \ref{fig2}\textit{A}.

Conceptualizing neurons as controllers in general and modeling them as DD-DCs in particular provides insights into multiple seemingly unrelated experimental observations. Firstly, it can explain the potentiation/depression transition in Spike-Timing Dependent Plasticity (STDP) and its asymmetry (Fig. \ref{fig1}{\it D})  \citep{markram1997regulation,bi1998synaptic,zhang1998critical}. Secondly, it can account for the temporal extent (non-instantaneous nature) of feedforward (STA) and feedback (spike-history dependent) filters and their adaptation to input statistics (Fig. \ref{fig2} {\it B-F}). Thirdly, it explains the loss of temporal precision in the neuronal spike-generation mechanism under constant input (Fig. \ref{fig3}, \textit{Right}) \citep{mainen1995reliability}. Fourthly, the operation of DD-DC in the online setting requires variability and/or noise, which is consistent with many neurophysiological observations \citep{rule2019causes,dobrunz1997heterogeneity,gordus2015feedback,olveczky2005vocal,deitch2021representational,schoonover2021representational,niell2010modulation}. Finally, viewing neurons as controllers is consistent with the observations of movement related activity throughout much of the brain including traditionally sensory areas \citep{musall2019single,zagha2022importance}.

Our model applies not only to neuroscience but also to machine learning and artificial intelligence. Current artificial neural networks are typically based on a neuronal unit inspired by an outdated view of neurons \citep{mcculloch1943logical,rosenblatt1961principles}. This neuronal unit is overly simplistic in that it lacks internal feedback and temporal dynamics (for more details, see below). Therefore, our proposed DD-DC model of a neuron could serve as an alternative foundational building block for constructing biologically-inspired artificial neural networks. 

\section*{The Direct Data-Driven Control (DD-DC) framework}

In this section we provide an overview of the DD-DC framework \citep{de2019formulas,willems2005note,markovsky2022data}. In our exposition, we use lowercase letters to denote scalar variables, lowercase boldface for column vector variables, and uppercase boldface for matrices and row vectors.

For the sake of clarity, we model the neuronal environment as a linear dynamical system in a discrete-time state-space representation (Fig. \ref{fig1}{\it A}):
\begin{align}
\label{dynamics}
\mathbf{x}_{t+1} & = \mathbf{Ax}_t + \mathbf{b}u_t, \\
\mathbf{y}_t & = \mathbf{Cx}_t,
\end{align}
where $\mathbf{x}_t\in\mathbb{R}^n$ represents the latent state of the environment at time $t$, $u_t$ is the control signal from the neuron, and $\mathbf{y}_t$ is the neuron's observation. In the realm of model-based control, the dynamics parameters $\mathbf{A}$, $\mathbf{b}$, and $\mathbf{C}$ are typically predefined, an unrealistic presumption in a biological context. We consider the system \eqref{dynamics} to be fully controllable and observable. In some cases, the optimal control signal is linearly related to the estimated latent state variable, $\mathbf{\hat{x}}$,
\begin{equation}
\label{control_law}
u_t=\mathbf{K} \mathbf{\hat{x}}_t.
\end{equation}

The DD-DC was developed for scenarios where the parameters $\bf A$, $\bf b$, and  $\bf C$ are unknown to the controller. It generates control signals directly from the observations, bypassing an explicit representation of the system dynamics and the latent state. This mapping is learned from historical pairings of observations and control signals. The intuition behind DD-DC is that any valid observation-control pairing belongs to the subspace delineated by \eqref{dynamics} spanned by historical pairings (Fig. \ref{fig1}{\it C}). This intuition is formalized by Willems' fundamental lemma \citep{willems2005note}, which posits that each observation-control pairing can be expressed as a linear combination of $k$ historical pairings ($1, 2, ..., k<t$),
\begin{equation}
\label{Willems}
\begin{bmatrix} \mathbf{\hat{x}}_{t+1}\\  \mathbf{\hat{x}}_t\\ u_t\\\end{bmatrix}=\begin{bmatrix} \mathbf{\hat{x}}_{1+1} & \hdots & \mathbf{\hat{x}}_{k+1} \\  \mathbf{\hat{x}}_1 & \hdots & \mathbf{\hat{x}}_k\\ u_1 & \hdots & u_k\\\end{bmatrix} \mathbf{g},  \;\;\;\;\;\; \mathbf{g}\in\mathbb{R}^k,
\end{equation}
assuming the matrix composed of $\bf \hat{x}$ and $u$ rows on the right has full row-rank (linear independence), a condition known as persistent excitation.

The DD-DC computes the control signal directly from historical pairings by solving \eqref{Willems} for $u_t$, thereby obviating the need for an explicit representation of the dynamical system ($\bf A$, $\bf b$, and $\bf C$) and  the latent state, ${\bf \hat{x}}$, in the controller. Initially formulated for ideal, noise-free, linear dynamics in offline settings with extensive datasets \citep{willems2005note}, DD-DC has recently been expanded to accommodate noisy observations, non-linear dynamics, online applications, and limited datasets \citep{coulson2019data,berberich2022linear,van2021beyond,rotulo2022online,markovsky2022data} making it a potent model of computation in biological neurons. In the following sections, we explore the implications of this hypothesis and demonstrate its alignment with existing experimental evidence and novel analysis.
 

\section*{Basic DD-DC Accounts for Spike-Timing-Dependent Plasticity (STDP)}

In this section, we demonstrate how even the most basic DD-DC neuron model can account for the principal characteristics of STDP (Fig. \ref{fig1}{\it D}). We start with the assumption that all dynamical variables are scalar ($n=1$) and that the system is fully observed, thus $\hat{x}=y=x$ (Fig. \ref{fig1}{\it B}). This allows us to express equation \eqref{dynamics} in a scalar form:

\begin{equation}
\label{scalar}
x_{t+1}=ax_t+bu_t.
\end{equation}

We posit that the neuron aims to stabilize the environment's state at $x=0$, even when its dynamics are unstable ($a>1$). To achieve this, we employ a one-step time-horizon Linear Quadratic Regulator (LQR), where the optimal control signal $u^*_t$ minimizes the sum of squared state error and control energy:
\begin{equation}
\label{lqr_cost}
u^*_t=\arg\min_{u_t} q \|x_{t+1}\|^2 + r \|u_t\|^2.
\end{equation}
This LQR objective is fulfilled by a linear control law:
\begin{equation}
\label{cl}
u^*_t=w^*x_t,
\end{equation}
with $w^*$, a scalar, representing the synaptic weight in place of $\mathbf{K}$ from \eqref{control_law}, as shown in Fig. \ref{fig1}{\it B}.

For simplicity, we initially address the limiting case of LQR with zero control cost ($r=0$), and later present the solution for nonzero $r$. In the $r=0$ scenario, \eqref{lqr_cost} is minimized by $x_{t+1}=0$. By substituting \eqref{cl} into \eqref{scalar} and ensuring $x_{t+1}=0$ for any given $x_t$ we deduce a closed-form LQR solution, $w^*=-a/b$.

Of course, neurons must implement this control law without prior knowledge of $a$ and $b$, a challenge adeptly addressed by the DD-DC model. Incorporating $x_{t+1}=0$ into \eqref{Willems}, we obtain:
\begin{equation}
\label{one-step-zero}
\begin{bmatrix} 0\\  x_t\\ u^*_t\\\end{bmatrix}=\begin{bmatrix} x_{1+1} & \hdots & x_{k+1} \\  x_1 & \hdots & x_k\\ u_1 & \hdots & u_k\\\end{bmatrix} \mathbf{g}=\begin{bmatrix} \mathbf{X}_+\\  \mathbf{X}\\ \mathbf{U}\\\end{bmatrix} \mathbf{g},
\end{equation}
where we introduced row-vector notation, ${\bf X}=\begin{bmatrix}x_1 & \hdots & x_k\end{bmatrix}$, ${\bf X}_+= \begin{bmatrix} x_{1+1} & \hdots & x_{k+1}\end{bmatrix}$, and ${\bf U}=\begin{bmatrix}u_1 & \hdots & u_k\end{bmatrix}$.

To determine the optimal control signal $u^*_t$, we first solve the top two rows of \eqref{one-step-zero} for $\mathbf{g}$. Given the underdetermined nature of the problem ($k$ typically exceeds the combined dimensions of $x$ and $u$), we express $\mathbf{g}$ via a pseudoinverse:
\begin{align*}
    \mathbf{g}= & \begin{bmatrix} \mathbf{X}_+^\top \; \mathbf{X}^\top \end{bmatrix} \left(\begin{bmatrix}
    \mathbf{X}_+ \\ \mathbf{X}\end{bmatrix}
\begin{bmatrix}\mathbf{X}_+^\top \mathbf{X}^\top \end{bmatrix}\right)^{-1}\begin{bmatrix}
    0\\ x_t \end{bmatrix}  = \\
    =  &  \frac{\begin{bmatrix} \mathbf{X}_+^\top \; \mathbf{X}^\top \end{bmatrix}}{\mathbf{X}_+ \mathbf{X}_+^\top\mathbf{X} \mathbf{X}^\top-(\mathbf{X}_+ \mathbf{X}^\top)^2} \begin{bmatrix}
    \mathbf{X}\mathbf{X}^\top  & -\mathbf{X}_+\mathbf{X}^\top\\ -\mathbf{X}\mathbf{X}_+^\top & \mathbf{X}_+\mathbf{X}_+^\top\end{bmatrix}\begin{bmatrix}
    0\\x_t \end{bmatrix}, & \numberthis \label{g}
\end{align*}

Subsequently, we substitute this $\mathbf{g}$ into the bottom row of \eqref{one-step-zero} to obtain 
\begin{align*}
u^*_t=\mathbf{U}\mathbf{g}=\frac{\mathbf{U}\mathbf{X}^\top \mathbf{X}_+\mathbf{X}_+^\top-\mathbf{U}\mathbf{X}_+^\top \mathbf{X}_+\mathbf{X}^\top}{\mathbf{X}_+\mathbf{X}_+^\top\mathbf{X}\mathbf{X}^\top -(\mathbf{X}\mathbf{X}_+^\top)^2}x_t, \numberthis 
\label{full_control}
\end{align*} 
This formulation can be interpreted as a control law \eqref{cl}, with 
\begin{align*}
w^*=\frac{\mathbf{U}\mathbf{X}^\top-\mathbf{U}\mathbf{X}_+^\top \mathbf{X}_+\mathbf{X}^\top(\mathbf{X}_+\mathbf{X}_+^\top)^{-1}}
{\mathbf{X}\mathbf{X}^\top-(\mathbf{X}\mathbf{X}_+^\top)^2 (\mathbf{X}_+\mathbf{X}_+^\top)^{-1}}.\numberthis \label{control_w}
\end{align*}
Notably, this control law obviates the need for a neuron to calculate $\mathbf{g}$ at each timestep or retain all past values of $\mathbf{U}$, $\mathbf{X}$, and $\mathbf{X_+}$. Instead, it requires only the storage and update of their covariances, a biologically plausible process previously utilized in similarity matching networks \citep{pehlevan2019neuroscience}. Rewriting these covariances as sums over recent history and omitting the denominator (a positive scalar indpendent of the control signal) yields:
\begin{equation}
\label{STDP}
w^*\sim \sum_{\tau=1}^k u_{\tau}x_{\tau} - \cos{(\widehat{\mathbf{X}\mathbf{X}_+})\sum_{\tau=1}^k u_{\tau}x_{\tau+1}},
\end{equation}
where $\cos{(\widehat{\mathbf{X}\mathbf{X}_+})}=(\mathbf{X}_+\mathbf{X}_+^\top)^{-1}\mathbf{X}_+\mathbf{X}^\top$. 

In a neurophysiological context, $x$ and $u$ in \eqref{STDP} symbolize pre- and post-synaptic neuronal activities, respectively, with non-zero values during spikes. Consequently, the sums in \eqref{STDP} accrue contributions solely when pre- and post-synaptic spikes are temporally proximate and depending on their temporal order. Note that although the indices of $u$ and $x$ are the same in the first sum, this is an artifact of the discrete-time setting, and $u$ must be delayed by at least a fraction of the time step to be computed from $x$. This model naturally accounts for the transition from potentiation to depression observed in STDP, as well as depression being weaker than potentiation (as $\cos{(\widehat{\mathbf{X}\mathbf{X}_+})}<1$ generically), aligning with empirical findings 
\citep{markram1997regulation,bi1998synaptic,zhang1998critical} (Fig. \ref{fig1}{\it D}).

The potentiation window of STDP can be viewed as the extension of the Hebbian rule \citep{hebb1949} to the temporal interplay between spikes. However, the rationale behind the relatively narrow depression window in STDP has remained elusive, largely due to its seemingly anti-causal nature. Current explanations of this phenomenon, e.g. \citep{gilson2011stability, aceituno2023learning}, rely on ad hoc assumptions. In contrast, our work shows that the apparent anti-causal aspect of STDP is a natural outcome of conceptualizing neurons as feedback controllers. Specifically, a pre-synaptic spike following a post-synaptic spike conveys information to the neuron about the effectiveness or ineffectiveness of its control of the environment. Thus, what is initially perceived as an anti-causal feature in STDP transforms into a causal mechanism when viewed through the lens of the feedback loop inherent in the controller model.

Next, we consider a DD-DC LQR controller with $r>0$, which in the scalar, one-step time-horizon  case is given by:
\begin{equation}
    w^*=\frac{ \mathbf{U} \mathbf{X}^\top \|\mathbf{X}_+\|^2 - \mathbf{U} \mathbf{X}_+^\top \mathbf{X} \mathbf{X}_+^\top}
{\|\mathbf{X}\|^2\|\mathbf{X}_+\|^2-(\mathbf{X} \mathbf{X}_+^\top)^2+ \left(\|\mathbf{U}\|^2\|\mathbf{X}\|^2- (\mathbf{U} \mathbf{X}^\top)^2\right)r/q}.
\label{sol_general}
\end{equation}
The full derivation of this solution is given in the Supplement but it aligns with the optimal LQR gain:
\begin{equation}
    w^*=\frac{-ab}{b^2+r/q}, 
\label{wopt}
\end{equation}
as confirmed by substituting $\mathbf{X}_+= a \mathbf{X} + b\mathbf{U}$ into \eqref{sol_general} and dividing both the numerator and the denominator of \eqref{sol_general} by the common expression $\left(\|\mathbf{U}\|^2\|\mathbf{X}\|^2- (\mathbf{U} \mathbf{X}^\top)^2\right)$ thus reducing it to \eqref{wopt}.

This solution provides a potential framework to interpret the variance in STDP profiles documented in various studies \citep{caporale2008spike} in terms of variance of $r$.

However, as our \eqref{STDP} involves covariances with only two time lags, it does not fully describe the time-course shown in Fig. \ref{fig1}{\it D}. This limitation, inherent to the scalar dynamics model \eqref{scalar}, motivates the exploration of higher order dynamics, as discussed in an upcoming section on reconstructing temporal filters.

\section*{Closed-loop DD-DC: malfunction under constant control law and restoration of function by adding noise to control}

\begin{figure}[hbtp]
     \centering

\includegraphics[width=0.47\textwidth]{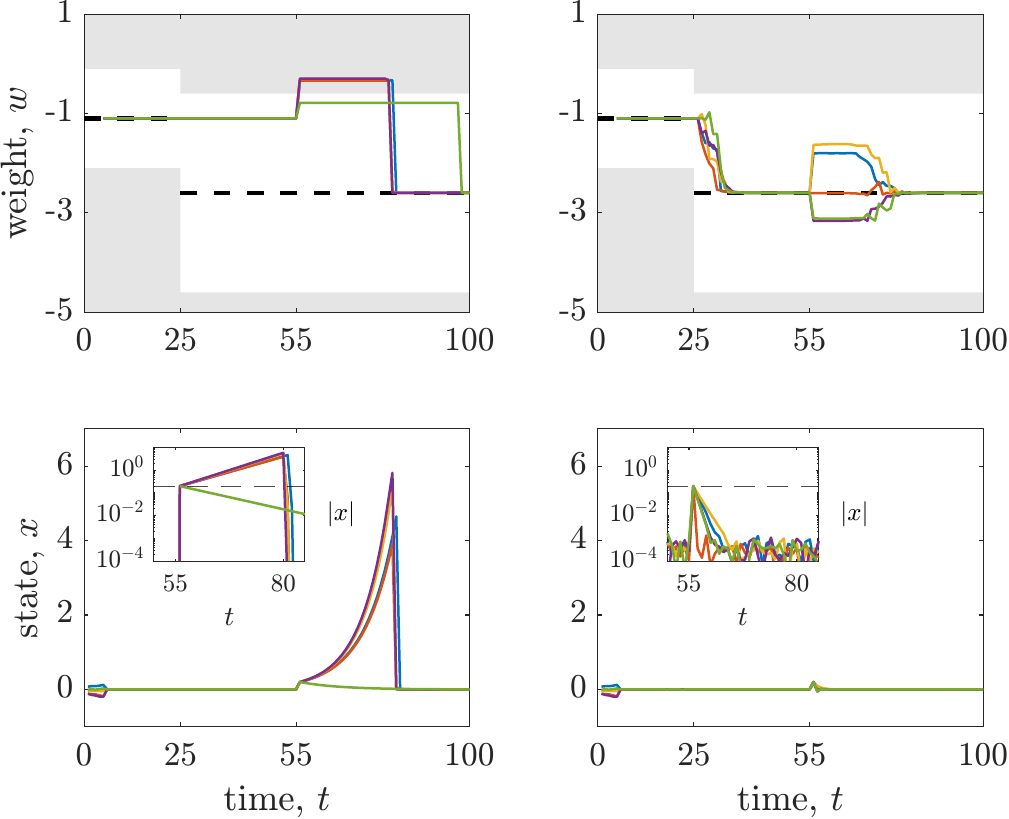}
        \caption[width=0.49\textwidth]
        {Dynamics of the weight, $w$, and state variable, $x$, in the online DD-DC LQR controller over time with the dynamical systems switch at time = 25. Closed loop control begins at $t=5$, at which point the weight, $w$, is constant and optimal (dashed line) and the state variable $x$ rapidly converges to zero. Post-switch, in scenarios without control noise (Left), the DD-DC controller struggles to adapt due to the loss of persistence of excitation, leading to a sub-optimal $w$; however, $x$ remains at zero, indicating no direct loss impact. To demonstrate the controller's maladaptation, a jolt $\Delta x = .2$ (dashed line in inset) is applied to the state variable $x$ at time $t= 55$. Introducing noise into the control law (Right), \eqref{cl_noise}, facilitates exploration thus re-establishing persistence of excitation and overall controller performance. This is evidenced by $x$ swiftly returning to zero and $w$ reverting to its optimal state. The white bands within the gray shaded areas in the $w$ plots represent regions of stability and instability respectively.}        
    \label{online}
\end{figure}

In this Section, we investigated the functioning of the DD-DC LQR controller through numerical simulations aimed at stabilizing a potentially unstable scalar dynamical system, \eqref{scalar}. Initially, we operated the controller in an open-loop mode for four time steps, implementing white-noise control, $u$, and tracking the resultant state variable, $x$. Subsequently, we computed the controller gain, $w$, by integrating the recorded values of $u$ and $x$ into the general LQR solution, \eqref{sol_general}. Following this initialization, we transitioned to a closed-loop operation of the DD-DC LQR controller, recalculating $w$ at each time step (see Supplementary Material for details). Our findings reveal that the DD-DC LQR controller successfully identifies and maintains the optimal value of $w$ up to time = 25, Fig. \ref{online} Left.

For the DD-DC controller to effectively replicate the adaptive behavior of a biological neuron, it must adjust to the evolving dynamics within a real-time, closed-loop framework. Accordingly, the update algorithm for $w$ incorporates a discount factor, progressively diminishing the influence of older data on covariance calculations (see Supplementary Material for details). Initially, the controller learns and applies the optimal value of $w$. However, when the parameters $a$ and $b$ undergo a switch (at time = 25), the controller fails to adapt, Fig. \ref{online} Left. To uncover the cause behind the DD-DC controller's failure to adjust following the static control law phase, we re-examined the data matrix entering \eqref{one-step-zero}: 
\begin{equation}
\label{degenerate}
\begin{bmatrix} x_{1+1} & \hdots & x_{k+1} \\  x_{1} & \hdots & x_{k}\\ u_{1} & \hdots & u_{k}\\\end{bmatrix}=\begin{bmatrix} x_{1+1} & \hdots & x_{k+1} \\  x_1 & \hdots & x_k\\ wx_1 & \hdots & wx_k\\\end{bmatrix}.
\end{equation}
Note that the sub-matrix composed of the $x$ and $u$ rows is rank deficient, thereby contravening the persistence of excitation condition. This rank deficiency signifies a critical limitation in the DD-DC's learning capability, as it impairs the system's ability to extract meaningful information from the data. Operationally, this issue manifests in the denominator of \eqref{sol_general} approaching zero.

Considering that sensory input may sometimes be constant \citep{pritchard1961stabilized} and control efforts typically aim for optimality, the question arises: how can the vulnerability of the online DD-DC controller be mitigated? Building upon the suggestions of control theorists \citep{rotulo2022online}, we propose that the brain deliberately generates variability and/or noise, denoted as $\eta$, to sustain the persistence of excitation condition even under static input or control regimes. Formally,
 \begin{equation}
 \label{cl_noise}
     u^*_t=w^*x_t+\eta_t,
 \end{equation}
Implementing the control law \eqref{cl_noise} in the same dynamical system is illustrated in Figure \ref{online} Right for VAR$(\eta_t)=10^{-6}$. The addition of low-variance noise to the control signal re-establishes the functionality of the DD-DC in an online setting. This noise reinstates the persistence of excitation, thereby restoring the DD-DC's operational efficacy. However, introducing too high-variance noise could cause significant deviations of the control signal from the LQR optimal \eqref{wopt}, implying that there might exist an ideal noise variance level for optimal control performance (see Supplementary Material for details).

Is there empirical evidence supporting the presence of such noise within the brain? Locally, such noise might originate from the inherent unreliability of synaptic transmission \citep{dobrunz1997heterogeneity}. On a broader scale, the brain encompasses specific circuits and neurons dedicated to introducing noise and/or variability, as evidenced in studies on songbirds \citep{olveczky2005vocal} and {\it C. elegans} \citep{gordus2015feedback}. Additionally, the brain can introduce variability in sensory inputs through the generation of corresponding motor outputs, exemplified by microsaccades \citep{ko2010microsaccades}. Given these mechanisms, the operational variability observed in neural representations \citep{deitch2021representational, schoonover2021representational}, appears less paradoxical and more a natural consequence of the brain's function as a DD-DC controller.

\section*{Reconstruction of feedforward and feedback temporal filters from data}

 \begin{figure*}[!h]
     \centering
     \includegraphics[width=\textwidth]
     {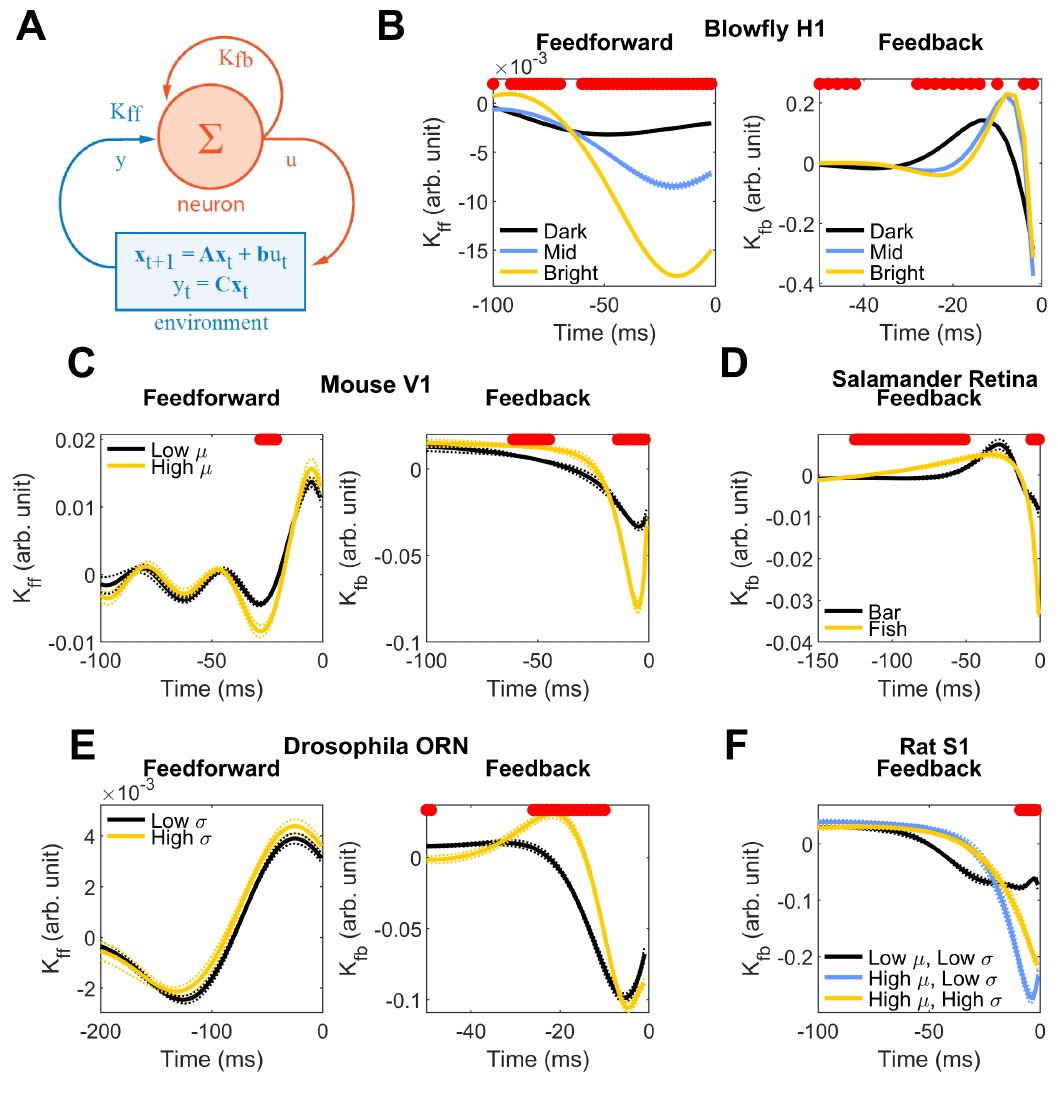}
\caption{{\bf A}. Illustration of a neuron modeled as an Auto-Regressive Moving Average (ARMA) controller, characterized by feedforward, ${\bf K}_{ff}$, and feedback, ${\bf K}_{fb}$, temporal filters. {\bf B-D}. Adaptation of experimentally measured temporal filters (depicted in black, yellow, and blue) to input signal statistics. Solid lines represent mean values, while thin dotted lines denote standard errors of the mean. Regions where differences are statistically significant (Wilcoxon rank-sum test with Bonferroni correction for multiple comparisons) are highlighted in red. {\bf B}. Variation in feedforward (akin to decorrelated Spike-Triggered Average, STA) and feedback (analogous to spike-history-dependence) filters of the blowfly H1 neuron \citep{lewen2001neural}, responding to visual motion against different background luminance levels. {\bf C}. Feedforward and feedback filters in pyramidal cells from mouse primary visual cortex \citep{teeter2018generalized} responding to current injections with varying mean levels. {\bf D}. Feedback filters in a salamander retinal ganglion cell \citep{palmer2015predictive} for stimuli comprising a drifting bar and a fish movie (feedforward filter data unavailable). {\bf E}. Adaptation of feedforward and feedback filters in a {\it Drosophila} Olfactory Receptor Neuron (ORN) \citep{gorur2017olfactory} to odorant concentrations with varying variances. {\bf F}. Feedback filters in rat somatosensory cortex pyramidal neurons \citep{rauch_lacamera_fusi2003noise}, responding to current injections modulated by an Ornstein-Uhlenbeck process atop a DC component. Feedforward filters are provided in the Supplement.}
    \label{fig2}
\end{figure*}
We now explore the DD-DC of a dynamical system of order $n > 1$, equipped with a scalar control signal, $u$, and a scalar observation, $y$ (Fig. \ref{fig2}\textit{A}). In such systems, observations are partial and insufficient for direct control, necessitating that the controller estimate the latent state, $\mathbf{x}$. For linear systems, this latent state can be inferred from recent sequences of observations and controls using time-delay embedding techniques  \citep{de2019formulas},
\begin{equation}
\label{tde}
\mathbf{\hat{x}}(t) = \begin{bmatrix} 
{y}_{t-n} \hdots {y}_{t-1}\;\;{u}_{t-n} \hdots {u}_{t-1} \end{bmatrix}^\top,\;\;\;\;\;
\end{equation}
enabling us to reformulate the control law, \eqref{control_law}, as:
\begin{equation}
{u}_t = \begin{bmatrix} \mathbf{K}_{ff} \; \mathbf{K}_{fb} \end{bmatrix} \begin{bmatrix} 
{y}_{t-n} \hdots {y}_{t-1}\;\;{u}_{t-n} \hdots {u}_{t-1} \end{bmatrix}^\top.
\end{equation}
Here, the feedforward, $\mathbf{K}_{ff}$, and feedback, $\mathbf{K}_{fb}$, temporal filters collectively form an Auto-Regressive Moving Average (ARMA) model of a neuron (Fig. \ref{fig2}\textit{A}).

In this Section, rather than optimizing feedforward and feedback temporal filters, we estimate them from experimental data. In our experiments, neurons are isolated from the loop and stimulated with sensory input or injected current, $y$, and the neuronal response, $u$, is recorded. These data are compiled into matrices $\mathbf{U}=\begin{bmatrix} {u}_1  & \hdots & {u}_t \end{bmatrix}$ and $\mathbf{\hat{X}} = \begin{bmatrix} \mathbf{\hat{x}}_1  & \hdots & \mathbf{\hat{x}}_t \end{bmatrix}$, which are linearly related:
\begin{equation}
\mathbf{U} = \begin{bmatrix} \mathbf{K}_{ff} \; \mathbf{K}_{fb} \end{bmatrix} \mathbf{\hat{X}}.
\end{equation}
We solve for the filters by linear regression using a pseudoinverse, 
\begin{equation}
\label{regression}
\begin{bmatrix} \mathbf{K}_{ff} \; \mathbf{K}_{fb} \end{bmatrix} = \mathbf{U} \mathbf{\hat{X}}^\top \left(\mathbf{\hat{X}} \mathbf{\hat{X}}^\top\right)^{-1}.
\end{equation}
Utilizing \eqref{regression}, we reconstruct the feed forward, ${\bf K}_{ff}$, and feedback, ${\bf K}_{fb}$, temporal filters from experimental data across various model systems, Fig. \ref{fig2} (see Supplement \ref{apdx:H1} for details). 

While these temporal filters have been previously measured \citep{rieke1999spikes,keat2001predicting,pillow2008spatio}, our controller-based perspective offers a novel interpretation. In a partially observed system, instantaneous observations alone are inadequate for control. Thus, feedforward and feedback filters with finite temporal extents are essential for control actions that are coherent with the latent state \eqref{tde}. The temporal extension of these filters aligns with the hypothesis that neuronal output influences the environment's latent state and acknowledges the partial observation of the system.

In each system we studied, neurons were recorded under various conditions, with stimulus statistics changing between these conditions. These included the blowfly H1 neuron with varying background luminance \citep{lewen2001neural}, mouse V1 pyramidal neurons responding to different mean injected current waveforms \citep{teeter2018generalized}, salamander retinal ganglion cells exposed to distinct visual stimuli \citep{palmer2015predictive}, {\it Drosophila} olfactory receptor neurons reacting to varying odorant concentrations \citep{gorur2017olfactory}, and pyramidal neurons in the rat somatosensory cortex stimulated with current injections of diverse means and variances \citep{rauch_lacamera_fusi2003noise}. As shown in Fig. \ref{fig2}, the filter shapes adapt to both the mean ($\mu$) and variance ($\sigma$) of the input statistics.

The adaptation of feedforward filters to input changes is well-documented \citep{srinivasan1982predictive,van1992theoretical,mainen1995reliability} and suggests a functional role beyond mere biological necessity, explainable by efficient coding and predictive information theories \citep{atick1992could,simoncelli2001natural,chalk2018toward}. Our analysis extends this understanding to feedback filters, which also adapt to changing stimulus statistics. This adaptation, not predicted by existing theories, calls for a new framework that treats feedforward and feedback filters equally, such as the controller neuron model.

\section*{Spike generation mechanism loses precision under constant input}

\begin{figure*}[hb]
     \centering
         \includegraphics[width=1\textwidth]
         {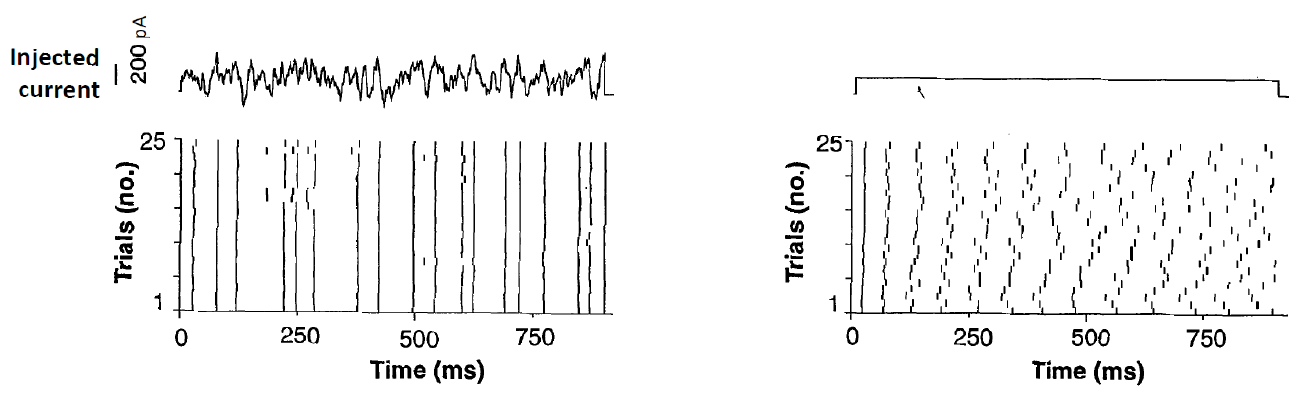}
\caption{Left: High-variance current injections into a neuron yield remarkably consistent spike trains over multiple trials, showcasing the precision of the spike-generation mechanism. Right: In contrast, a constant current input leads to notably variable spike trains, revealing a significant reduction in spike-timing precision. This dichotomy highlights the neuron's differential response to varying and constant stimuli \citep{mainen1995reliability}.}
    \label{fig3}
\end{figure*}

Neuronal spike generation typically showcases remarkable precision: repeated injections of the same current waveform into a neuron yield highly reproducible spike trains, precise down to milliseconds \citep{bryant1976spike, mainen1995reliability} (Fig. \ref{fig3}, \textit{Left}). This level of precision in spike timing must incur metabolic cost and is therefore suggestive of a functional significance. Intriguingly, this precision deteriorates when the neuron is subject to a constant current input \citep{mainen1995reliability} (Fig. \ref{fig3}, \textit{Right}), exposing a notable limitation in the spike-generation mechanism.
The DD-DC model of neuronal function offers an insightful explanation for this observed decline in spike-timing precision with constant input. The DD-DC model posits that a neuron reconstructs any state as a weighted sum of past states, which is effective only when these past states are sufficiently varied \citep{willems2005note}. This is rooted in the persistency of excitation condition, requiring the matrix of past states in \eqref{Willems} to have full row-rank. Under constant input, however, this condition fails as the lag vectors in \eqref{tde} become uniform. Consequently, when a neuron processes recent history (approximately 100ms in Fig. \ref{fig3}, \textit{Right}), the DD-DC model predicts erratic outputs in response to a constant current, mirroring the vulnerability of the spike-generation mechanism to such inputs. 

Traditionally, the variability in spike timing under constant input was ascribed to intrinsic ion channel noise, not controlled for in experimental setups \citep{schneidman1998ion}. This noise was thought to be inconsequential for spike timing in the presence of highly variable inputs, as the effects would be overshadowed by the abundance of open ion channels. However, even slight variations in injected current (STD $\approx 50$pA) are observed to restore spike-timing precision \citep{rauch_lacamera_fusi2003noise}, posing questions about the underlying mechanisms of such sensitivity. Our model offers a different perspective, suggesting that this sensitivity to low-variance noise stems from the high condition number (ratio of the largest to the smallest singular values) of time-delay covariance matrices. These matrices must be inverted to compute a neuron's response (akin to \eqref{regression}). Our hypothesis posits that the singularity at constant current can be empirically validated by measuring spike time variability against noise variance below the threshold reported in \citep{mainen1995reliability}, and correlating it with the condition number of the time-delay covariance.

\section*{Discussion}

The power of the proposed DD-DC model of a neuron is in that, starting from a single postulate, it offers explanations for multiple seemingly unrelated neurophysiological phenomena, including the switch between potentiation and depression in STDP and its asymmetry, the extended nature and input-dependent adaptation of feedforward and feedback temporal filters, the imprecision of the spike-generation mechanism under constant input, and the prevalence of operational variability and noise in the brain. Although each of these explanations provides only a circumstantial evidence, their multitude and variety provide strong support for the DD-DC model. This perspective has the potential to deepen and refine our understanding of the brain, and may also aid in the development of biologically-inspired artificial neural networks.

\begin{figure*}[b]
     \centering
    \includegraphics[width=0.8\textwidth]{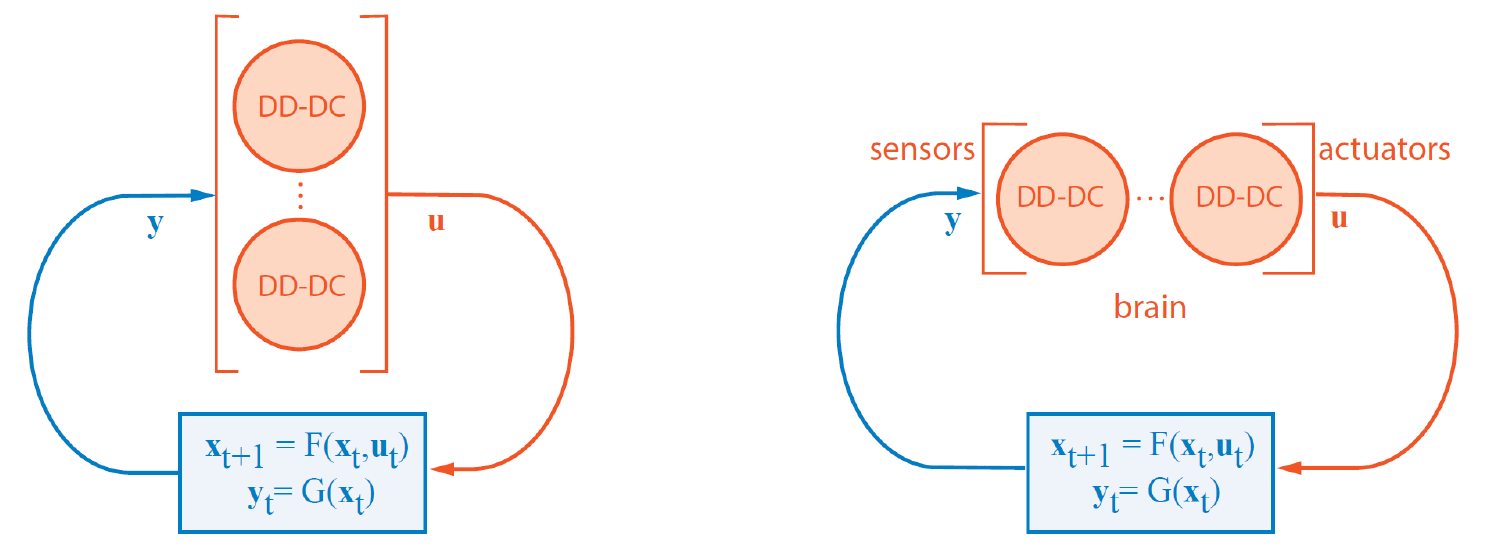}
\caption{Left: Illustration of controlling a nonlinear dynamical system using multiple switching DD-DCs. Right: Depiction of a deep network model where each neuron exerts control over its immediate environment, contributing to the broader control exerted by the entire brain over the external environment.}
    \label{fig4}
\end{figure*}

{\bf Nonlinear dynamics and control.} In this study, we focused on a DD-DC model that assumes discrete-time and linear dynamics of the environment. In reality, the dynamics are continuous-time and nonlinear. As is often the case in control theory, we expect that our framework can be naturally extended to continuous time. As nonlinear dynamics of the loop can be approximated locally as linear, we speculate that they can be modeled by a switching linear system controlled by a set of switching DD-DCs (see Fig. \ref{fig4}, \textit{Left}). This would explain why layers of processing in the brain contain many neurons in parallel performing analogous functions. 

How to derive a nonlinear controller model of a neuron from the normative perspective? Even for linear plant dynamics, apart from special cases like LQR, the optimal controller may not be linear. Perhaps, neuronal action potentials, in addition to having higher information transmission capacity relative to graded potentials in noisy environments
\citep{jack1975electric,sarpeshkar1998analog,laughlin1998metabolic}, have other operational advantages similar to widely used bang-bang control \citep{kamien2012dynamic}. Deriving such a controller may help model neurons with active conductances and spikes \citep{koch2004biophysics} on the algorithmic level.

{\bf Stability and performance objectives.} In control theory, the stability of the closed loop is of primary concern. While a stability criterion for DD-DC can be formulated  \citep{de2019formulas}, it allows for numerous solutions. How to select a specific solution out of the stable set is not clear. One approach could be to look for the most stable solution, which would remain stable even in the presence of noise and uncertainty about the parameters of the dynamical system. Another approach could be based on the observation that the brain operates at the edge of chaos  \citep{maass2002real,beggs2003neuronal}, favoring borderline stable solutions. A similar borderline stable solution is suggested by viewing a neuron as an integrator, which would require the top eigenvalue to have a unit norm \citep{seung1996brain}. Such flexibility in the choice of the objective may allow one to use different solutions to model different neuronal classes.

{\bf A network of DD-DC neurons.} The DD-DC model of a neuron presented here lumps the rest of the neurons into a single dynamical system, yet each fellow neuron can also be modeled as a DD-DC. This raises the question of how multiple DD-DC neurons interact with each other in a network. We leave this question to future work and comment only on several experimental observations that support this view.

First, measurements of synaptic plasticity may shed light on the temporal lag caused by the feedback loop. In the case of STDP (Fig. \ref{fig1}, \textit{Right}), pre- and post-synaptic spikes must be almost synchronous for plasticity to occur, indicating that the loop traverses an order of one synapse. This finding corresponds to the known abundance of short local feedback loops in the cortex \citep{zilberter2000dendritic,song2005highly}. At the other extreme, the plasticity of some synapses in the cerebellum and the hippocampus peaks when the spikes lag by tens of milliseconds  \citep{dudman2007role, suvrathan2016timing}. This suggests longer loops involving different brain regions or even the external environment \citep{Jayabal2022.11.28.518128}. The abundance of loops is not limited to mammalian brains \citep{bell1999levels} and has been reported in invertebrates as well \citep{varshney2011structural,winding2023connectome}.

Second, the ability of individual neurons to control long (multi-synaptic and trans-environment) loops may seem unrealistic. However, theoretical analysis \citep{london2010sensitivity, monteforte2012dynamic} and experimental observations seem to support long-range propagation of signals from individual neurons \citep{wolfe2010sparse}. Specifically, rodents can be trained to behaviorally report single-neuron electrical stimulation in the barrel cortex \citep{houweling2008behavioural} suggesting that the spikes of a single neuron make impact sufficiently far downstream to elicit behavior. Also, stimulation of single neurons in the motor cortex evoke whisker movements \citep{brecht2004whisker} suggesting that single neurons can produce observable effects on the environment. Taken together, these experiments support the propagation of individual neurons' spikes around long loops.

Third, modeling neurons as DD-DC controllers offers an explanation for the representation of movement outside of the motor cortex \citep{musall2019single,zagha2022importance}.

As DD-DC neurons in every layer (Fig. \ref{fig4}, \textit{Right}) combine systems identification and control, they acquire characteristics of both sensory and motor representations. Therefore, it is natural to expect a mixed representation at every layer with a gradual transformation from sensory to motor. The controller perspective also accounts for the context dependence of neuronal representations \citep{niell2010modulation}. As neuronal activity should not just reflect the sensory stimulus but rather optimal control, which is context dependent, it is natural to expect such representations.

\section*{Relationship to Other Work}

The concept of modeling neurons as controllers intersects with several established research avenues. A notable idea in neuroscience relates neuronal output to a prediction of future inputs \citep{srinivasan1982predictive,tishby2000information,rao1999predictive,chalk2018toward}. To optimize control the DD-DC neuron implicitly infers environmental dynamics, which could also be used for prediction. However, the controller neuron does not just predict the future input but aims to influence it through its output.

Normalization models, which have been both experimentally observed and theoretically justified \citep{wainwright200210,carandini2012normalization,brenner2000adaptive}, resemble the feedback filter in our model. However, these models focus on interactions among parallel channels under static stimuli, overlooking temporal correlations and stimulus dynamics. In contrast, the DD-DC model proactively controls inputs through its influence on the underlying dynamical system.

Our approach also aligns with the idea that spiking neuron networks encode temporally variable inputs \citep{boerlin2013predictive}. Although these networks are predominantly feedforward and don't allow neuron outputs to modify network inputs, both concepts emphasize learning generative dynamics.

Data-driven control in network contexts, including brain networks, has been investigated \citep{baggio2021data}. However, these studies generally involve controlling networks through external perturbations and lack a focus on biological plausibility. Unlike our neuron-centric DD-DC model, they require access to multiple network nodes and are constrained by the resolution of technologies like fMRI.

The DD-DC ARMA model for neurons (Fig. \ref{fig2}\textit{A}) shares similarities with Generalized Linear Models (GLMs) \citep{pillow2008spatio}, notably in possessing feedforward and feedback filters. But, while GLMs are stochastic and nonlinear, the DD-DC model is deterministic, linear, and provides a rationale for the duration and adaptation of these filters. The concept of temporal integration in our model also echoes the principles of integrate-and-fire models \citep{lapicque1907recherches}, laying groundwork for future connections between these theories.

Differing from the conventional McCulloch-Pitts-Rosenblatt unit in artificial neural networks, the DD-DC neuron integrates inputs over time and features an auto-regressive loop, unlike the instantaneous response of standard units. Also, in contrast to network-wide optimization in artificial neural networks, the DD-DC model optimizes objectives at the neuronal level.

Neurons are sometimes conceptualized as agents in the reinforcement learning (RL) paradigm, \citep{sutton2018reinforcement,dayan2002matters}. While control theory and RL share commonalities, key distinctions include control theory's implicit dynamical systems model of the environment and its focus on optimizing specific objectives based on controls and observations, as opposed to the reward-maximization approach in RL.

Previous studies have conceptualized the whole brain as a controller acting on the external world \citep{cisek1999beyond,cisek2010neural,ahissar2016perception}, and used LQR with delays/noise to model internal feedback \citep{li2022internal}. Our DD-DC approach extends this concept to individual neurons. If both the entire brain and single neurons can be modeled as controllers, intermediate levels of brain structure might also fit this model \citep{robinson1981use,miles1981plasticity}. Earlier models separated sensory system identification and motor control \citep{haruno2001mosaic}, but our unified neuron-as-controller model eliminates the need to match corresponding sensory and motor units. Future research could explore how coherent controller actions emerge in self-organized networks of such neuron-controllers.

\acknow{We would like to express our sincere gratitude to Anirvan Sengupta, Luca Mazzucato, Christian Pehle, and Subham Dey for their invaluable discussions and insights. Our special thanks to Lucy Reading-Ikkanda for her exceptional assistance with the illustrations. Additionally, we acknowledge the use of ChatGPT in the writing process.}

\bibliography{main}

\pagebreak

\section*{Supplementary Information}

\section*{A. Derivation of the DD-DC LQR controller }

\begin{figure}[ht]
  \centering
  \includegraphics[trim=100 220 200 20,clip,width=.49\textwidth]{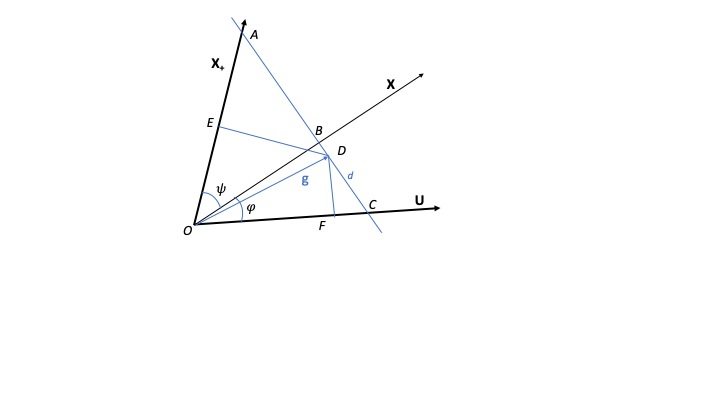}
  \caption{Illustration for the derivation of DD-DC LQR controller. \label{lqrgeom}}
\end{figure}

Given an initial state $x_0$, controller finds a control signal $u^*$ that brings the state to the value $x^*$ minimizing  LQR loss
\begin{equation}
       \|x^*\|^2 + r/q \|u^*\|^2.
 \end{equation}
Corresponding DD-DC optimization problem takes the form:
\begin{align}
    \min_{\mathbf{g}}  &(\mathbf{g}\mathbf{X}_+^\top)^2 + r/q(\mathbf{g}\mathbf{U}^\top)^2 \\
    \text{s.t.   } &(\mathbf{g}\mathbf{X}^\top)=x_0
 \end{align}
The chart in Fig. \ref{lqrgeom} 
shows sample vectors $\mathbf{X},\mathbf{U},\mathbf{X}_+$ which are co-planar due to the dynamics. Without loss of generality, We consider the solution vector $\mathbf{g}$ to be co-planar with them. The blue line $AC$ orthogonal to $\mathbf{X}$ is the locus of solutions satisfying the constraint, so that $|OB|=x_0/\|\mathbf{X}\|$. We see that $|OC|=|OB|/\cos\phi$, $|OA|=|OB|/\cos\psi$, $|BC|=|OB|\tan\phi$, $|BA|=|OB|\tan\psi$. 
Denote $d:=|DC|$, we'll be solving the problem in this variable. Let $DF, DE$ be orthogonal to $\mathbf{U},\mathbf{X_+}$, then observe that $\angle CDF=\phi,\angle ADE=\psi$ as angles with orthogonal sides. This gives expressions:
\begin{align*}
    |OF|&=|OB|/\cos\phi -d\sin\phi,\\
    |OE|&=|OB|/\cos\psi -|OB|(\tan\phi+\tan\psi)\sin\psi-d\sin\psi.
\end{align*}
Note also that $\mathbf{g}\mathbf{X}_+^\top=|OE|\|\mathbf{X}_+\|$,  $\mathbf{g}\mathbf{U}^\top=|OF|\|\mathbf{U}\|$. Substituting all that, we can rewrite the objective as quadratic expression in $d$. Solution takes the form
\begin{align*}
    d = \frac{ (-\sin\psi\cos\psi + \sin^2\psi\tan\phi)\|\mathbf{X_+}\|^2 +\tan\phi\|\mathbf{U}\|^2 r/q}{\cos\phi\cos\psi(\sin^2\psi\|\mathbf{X_+}\|^2 + \sin^2\phi\|\mathbf{U}\|^2 r/q)} \frac{x_0}{\|\mathbf{X}\|} 
\end{align*}
The optimal control signal we seek is $u^*=\mathbf{g}\mathbf{U}^\top=|OF|\|\mathbf{U}\|$, which gives:
\begin{align*}
    u^* &= \frac{\cos\phi\sin\phi  + \cos\psi\sin\psi -\cos\psi\sin\psi\sin^2\phi -\cos\phi\sin\phi\sin^2\psi  }{\cos\phi\cos\psi(\sin^2\psi\|\mathbf{X_+}\|^2 + \sin^2\phi\|\mathbf{U}\|^2 r/q)} \\ &\times \frac{ x_0\|\mathbf{U}\|\|\mathbf{X_+}\|^2 \sin\psi}{\|\mathbf{X}\|}
\end{align*}
For online calculation it is convenient to have an expression in variances and covariances of the variables, so we reformulate trigonometric expressions through cosines only. For that purpose we group terms first and fourth, second and third from the numerator of the first fraction:
\begin{align*}
    &\cos\phi\sin\phi(1-\sin^2\psi)  + \cos\psi\sin\psi(1-\sin^2\phi) \\
    &=\cos\phi\sin\phi\cos^2\psi  + \cos\psi\sin\psi\cos^2\phi) \\ &=
\cos\phi\cos\psi\sin(\phi+\psi) 
\end{align*}
Bringing in $\sin\psi$ from the second fraction numerator we use cosine of differences rule to observe:
\begin{align*}
    \cos\phi\cos\psi\left[ \sin(\phi+\psi)\sin\psi \right] = \cos\phi\cos\psi\left[\cos\phi - \cos\psi\cos(\phi+\psi) \right]
\end{align*}
Now we can rewrite the expression for $u^*$:
\begin{align*}
    u^* = \frac{\cos\phi - \cos\psi\cos(\phi+\psi) }{(1-\cos^2\psi)\|\mathbf{X_+}\|^2 + (1-\cos^2\phi)\|\mathbf{U}\|^2 r/q}  \frac{ \|\mathbf{U}\|\|\mathbf{X_+}\|^2} {\|\mathbf{X}\|}x_0
\end{align*}
Replacing $\cos\phi=\tfrac{\|\mathbf{X}\mathbf{U}^\top\| }{\|\mathbf{X}\|\|\mathbf{U}\|}$, $\cos\psi=\tfrac{\|\mathbf{X}\mathbf{X_+}^\top\| }{\|\mathbf{X}\|\|\mathbf{X_+}\|}$, and $ \cos(\phi+\psi)=\tfrac{\|\mathbf{X_+}\mathbf{U}^\top\| }{\|\mathbf{X_+}\|\|\mathbf{U}\|}$, we obtain $u^*=w^* x_0$ where $w^*$ is defined in \eqref{sol_general}.

\section*{B. Simulation of the DD-DC LQR controller }

The simulation presented in Fig. \ref{online} begins in the open loop mode with random initialization of $x$ and random Gaussian control signal with standard deviation 0.01. Open loop runs for 4 steps, after that the control weight, $w$, is estimated using \eqref{sol_general} and then updated at each time step in the closed loop setting using the variances and covariances of the variables $x,u,x_+$. In the online setting, these variances and covariances are updated at each time step according to
\begin{equation*}
    \text{cov}_{\alpha,\beta}(t+1) = \gamma\text{cov}_{\alpha,\beta}(t) +(1-\gamma)\alpha(t)\beta(t),
\end{equation*}
where $\alpha, \beta$ are any of $x,u,x_+$. For this simulation we set $\gamma=0.5$. For the noise process,  Fig. \ref{online} Right, Gaussian noise with standard deviation $\sigma=0.001$ was added to the control signal $u$.

To check the ability of the solution to adapt to changes of the control system parameters $a, b$ were abruptly switched at steps 25 as follows:  $a=1.1, 1.3;
b=1, 0.5$. In addition, a jolt was added to the system state $x$: $0.2$ at step 55. Throughout the simulation, the LQR penalties were kept constant at $r/q=0$.

\begin{figure}[hbtp]
     \includegraphics[width=0.47\textwidth,right]{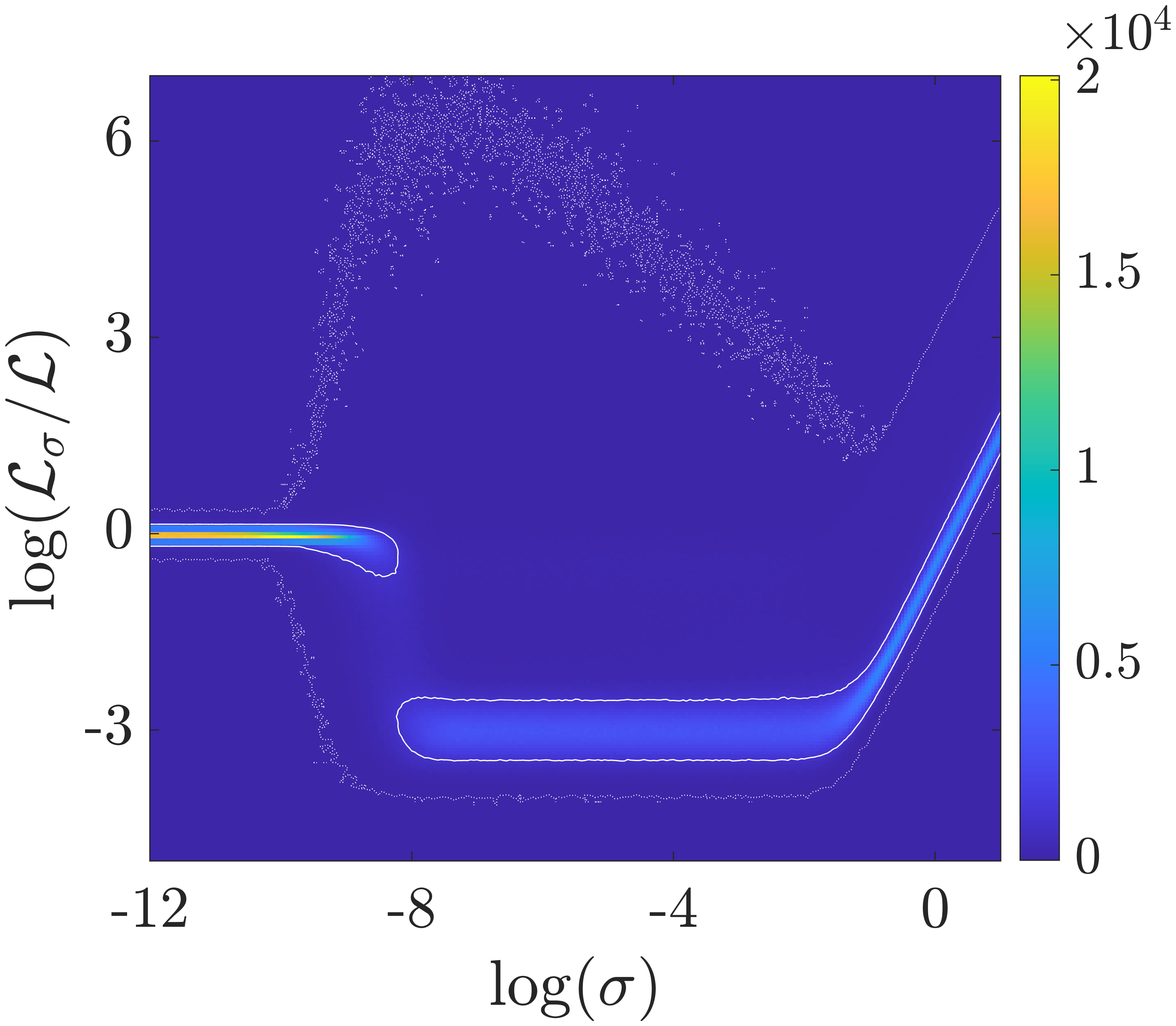}
        \caption[width=0.49\textwidth]
        {Heat map of the ratio of losses $\mathcal{L}_\sigma=\sum_t x_t^2(\sigma)$ for different levels of Gaussian noise $\sigma$ over the no-noise loss $\mathcal{L}=\sum_t x_t^2(0)$.
         For each noise level, $\sim 5\times 10^4$ simulations were run ($\sim 10^7$ trials total). The dotted line encloses all trials, and the solid line encloses 95\% of all observed trials. For the majority of trials the ratio is less than one indicating that the addition of noise is advantageous. The average loss ratio is observed to be lowest for $\sigma$ in the range $10^{-8}-10^{-2}$.}        
    \label{lossHist}
\end{figure}

\section*{C. Datasets}\label{apdx:H1}

\subsection*{C.1 Blowfly neuron H1 (Fig. \ref{fig2}B in the main text)}
Flies maintain their flight heading by utilizing visual feedback to modulate wingbeat-amplitude asymmetry. The bilaterally symmetric pair of H1 neurons (one for each direction of motion), processes information for horizontal direction (yaw) control \citep{rieke1999spikes}. The firing of these neurons not only signals the horizontal velocity of optic flow \citep{rieke1999spikes} but also plays a role in controlling wingbeat asymmetry \citep{williams2018blowfly}. In the experiment a computer controlled stepper motor was used to rotate the fly along a vertical axis while spike trains from H1 were recorded extracellularly. The fly sat immobilized in wax in a plexiglass cylinder to allow temperature to be controlled at 22 degrees Celsius. 

The setup was situated outdoors in a wooded area on a bright day \citep{lewen2001neural}. The experiment started 42 minutes before sunset and ended 48 minutes after sunset. During this 90 minute interval, light intensities decreased by a factor of approximately $6 \times 10^5$. The fly was subjected to a repeated yaw motion (period 10 seconds), presented 540 times. The yaw velocity was synthesized by a Markov model derived from a video recording of flies in pursuit flight, and is therefore representative of natural flight. Regression was performed separately for 4 parts of the 540 sessions: two during bright light time, one - middle, one - dark. The two bright parts were very similar, so filters are presented for the 1st, 3rd, and 4th part.

\subsection*{C.2 Mouse V1 pyramidal neuron (Fig. \ref{fig2}C in the main text)}
Response profiles of pyramidal neurons in primary visual cortex were measured using patch clamp recordings in brain slices from adult mice between the ages of P45 to P70 \citep{teeter2018generalized}. The specific neuron analyzed here was from a fluorescently-labeled neuron from a mouse expressing tdTomato in a Cre-dependent manner in a Cux2-Cre transgenic mouse line, which labels excitatory neurons in cortical layers 2/3 and 4. 

Voltage was recorded while injecting a stimulus composed of pink noise riding on top of square wave pulses. Pulses were 3 seconds long at rheobase (“Low $\mu$”) or 1.5 times rheobase (“High $\mu$”). Pink noise with a coefficient of variation equal to 0.2 was generated using 2 different seeds. Each stimulus was applied to the neuron 4 times for a total of 8 trials of Low and High $\mu$. Recordings were initially performed at a sampling rate of 200 kHz. Action potentials were detected using standard approaches and the binary spiking activity was downsampled to 1 kHz for analysis.

\subsection*{C.3 Salamander retinal ganglion cell (Fig. \ref{fig2}D in the main text)}
In response to the visual stimulation, retinal ganglion cells generate reproducible spike trains which can be predicted using a combination of feedforward and feedback temporal filters \citep{keat2001predicting, pillow2008spatio}. Retinal ganglion cells from larval tiger salamander retina were extracellularly recorded from freshly dissected retina pressed onto a multi-electrode array. 

Movies were presented to the retina at 60 frames/second and voltages were recorded at 10 kHz. The “Bar” stimulus consisted of a black bar moving against a gray background according to the Brownian motion of a particle bound by a spring to the center of the display. This was repeated 62 times, with each trial lasting ~8 seconds. The “Fish” stimulus consisted of a ~18-second clip of a fish swimming in a tank with swaying plants in the background. This stimulus was repeated 102 times. Action potentials were detected using standard approaches and the binary spiking activity was downsampled to 1 kHz for analysis.

\subsection*{C.4 Drosophila olfactory receptor neuron (Fig. \ref{fig2}E in the main text)}
Responses of olfactory receptor neurons (ORN) in Drosophila was studied by stimulating them with odors of varying mean concentration (bias) and variance of the stochastically varying component \citep{gorur2017olfactory}.  Neurons were recorded extracellularly from the ab3 sensillum of Drosophila. The neuron analyzed here was an ab3A ORN. 

A stimulus of fluctuating ethyl acetate was delivered to the fly antenna by blowing air over pure monomolecular odorants in liquid phase. The gas concentration was determined by the flow rate of air. For the recordings analyzed here, the variance of the signal changed every 5 seconds around a constant mean, with the variance switching a total of 60 times for 30 trials of “High $\sigma$” and 30 trials of “Low $\sigma$” stimulation. These stimuli were Gaussian distributed, generated by optimizing control signals to Mass Flow Controllers which regulated airflows. The mean stimulus corresponded to an odorant flux of approximately 2.5 $\mu$ mol/s. The “High $\sigma$” condition corresponded to a standard deviation of approximately 0.75 $\mu$ mol/s, while the “Low $\sigma$” condition corresponded to a standard deviation of approximately 0.3 $\mu$ mol/s. Identified action potentials were binarized and downsampled to 1 kHz for analysis. Odorant concentrations were also recorded at 1 kHz.

\subsection*{C.5 Rat somatosensory pyramidal neuron (Fig. \ref{fig2}F in the main text)}
Response profiles of pyramidal neurons in rat somatosensory cortex were measured using patch clamp recordings in brain slices \citep{rauch_lacamera_fusi2003noise}. The neuron analyzed here was recorded in current-clamp whole cell configuration from the soma of a layer 5 regular spiking pyramidal cell. Data were recorded at 5 kHz and low-pass filtered at 2.5 kHz.

Current was modeled by an Ornstein-Uhlenbeck process with different resulting means $\mu$\ and standard deviations $\sigma$\. Low $\mu$\ corresponded to a mean of approximately 700 pA. Low $\sigma$\ corresponded to a standard deviation of approximately 200 pA. High $\mu$\ corresponded to approximately 1800 pA. High $\sigma$\ corresponded to 300-400 pA.

\section*{D. Filter Fitting Procedure}
\subsection*{D.1 Data Preparation}
First, spiking vector $\mathbf{U}=[u(1)\dots u(T)]$ and stimulus vector $\mathbf{Y}=[y(1)\dots y(T)]$ were z-scored across time to have 0 mean and unit variance.
Lag matrices were then constructed by stacking $n$-dimensional lag vectors together for the whole length of the dataset, $T$:
\[
\mathbf{U}_{\text{lag}} = \begin{bmatrix}
     u(1) & u(2) & \ldots & u(T-n+1) \\
    u(2) & u(3) & \ldots & u(T-n+2) \\
    \vdots & \vdots & \ddots & \vdots  \\
    u(n) & u(n+1) & \ldots & u(T)
\end{bmatrix}
\]
where $\mathbf{U}_{\text{lag}}$ has dimensions $n \times (T-n+1)$. Similarly:

\[
\mathbf{Y}_{\text{lag}} = \begin{bmatrix}
    y(1) & y(2) & \ldots & y(T-n+1) \\
    y(2) & y(3) & \ldots & y(T-n+2) \\
    \vdots & \vdots & \ddots & \vdots  \\
    y(n) & y(n+1) & \ldots & y(T)
\end{bmatrix}
\]
where $\mathbf{Y}_{\text{lag}}$ has dimensions $n \times (T-n+1)$.

\subsection*{D.2 Dimensionality Reduction}
To avoid overfitting and promote smoothness of the filters, we projected the matrices $\mathbf{U}_{\text{lag}}$ and $\mathbf{Y}_{\text{lag}}$ onto a low-dimensional subspace. The stimulus matrix $\mathbf{Y}_{\text{lag}}$ was projected onto the top $N$ principal components, where $N$ is the number of principal components needed to capture 75\% of the variance of the data. If the singular value decomposition of $\mathbf{Y}_{\text{lag}} = \mathbf{W\Sigma V}^\top$,
then define $\mathbf{S} = \mathbf{\Sigma V^\top}$, and define $\mathbf{Y_{\text{PCA}}}$ as the first $N$ rows of $\mathbf{S}$, i.e., $\mathbf{Y_{\text{PCA}}}$  has dimensions $N \times (T-n+1)$.

The spiking matrix $\mathbf{U}_{\text{lag}}$ was dimensionally reduced using Laguerre polynomials, a family of orthogonal polynomials under the exponential kernel. The first few Laguerre polynomials, $L_l$, are: 
\vspace{-0.1in}
\begin{table}[htbp]
\centering
\begin{tabular}{cc}
\hline
$l$ & $L_l(z)$ \\
\hline
0 & 1 \\
1 & $-z + 1$ \\
2 & $\frac{1}{2}(z^2 - 4z + 2)$ \\
3 & $\frac{1}{6}(-z^3 + 9z^2 - 18z + 6)$ \\
4 & $\frac{1}{24}(z^4 - 16z^3 + 72z^2 - 96z + 24)$ \\
5 & $\frac{1}{120}(-z^5 + 25z^4 - 200z^3 + 600z^2 - 600z + 120)$ \\
\hline
\end{tabular}
\end{table}

To define a natural time course for the filters and to ensure orthogonality, the Laguerre polynomials were each multiplied by a decaying exponential. Also a timescale $\tau$ was introduced, producing functions $\Lambda_{l,\tau}(x)=L_l(x/\tau)\exp(-x/2\tau)$. These functions were calculated on values of $x=0,1\dots n-1$. For any given timescale $\tau$ and number of polynomials $p$ we assembled a $n\times p$ matrix $\mathbf{\Lambda}$, where $\mathbf{\Lambda}_{i,j}=\Lambda_{i,\tau}(j-1)$. We can now define $\mathbf{U_{\text{LGR}}} = \mathbf{\Lambda}^\dagger \mathbf{U}_{\text{lag}}$, where $\mathbf{U_{\text{LGR}}}$ has dimensions $p \times (T-n+1)$ and $\dagger$ signifies Moore-Penrose pseudo-inverse.
\subsection*{D.3 Computing Filters}
Regressor weights were then calculated as $[ \mathbf{\tilde{K}}_{ff} \ \mathbf{\tilde{K}}_{fb}] = \mathbf{U}\mathbf{\tilde{X}}^\top(\mathbf{\tilde{X}} \mathbf{\tilde{X}}^\top)^{-1} $, where $\mathbf{\tilde{X}} = [ \mathbf{Y_{\text{PCA}}^\top} \ \mathbf{U_{\text{LGR}}^\top]^\top}$, and $\mathbf{\tilde{X}}$ has dimensions $(N+p) \times (T-n+1)$, $\mathbf{\tilde{K}}_{ff}$ has dimensions $1 \times N$, and $\mathbf{\tilde{K}}_{fb}$ has dimensions $1 \times p$.
Then the time-dependent kernels were calculated as:
\[
\begin{aligned}
    \mathbf{K}_{ff} & = \mathbf{\tilde{K}}_{ff}\mathbf{W}(1:N, :) \\
    \mathbf{K}_{fb} & = \mathbf{\tilde{K}}_{fb} \mathbf{\Lambda}^\dagger
\end{aligned}
\]
\subsection*{D.4 Parameter Search}
The number of Laguerre polynomials, $p$, and the timescale, $\tau$, were chosen to minimize the reconstruction error of the fit using a grid search, with the number of polynomials restricted between 2 and 7 and $\tau$ chosen from $\{1, 2, 4, 8, 16, 32, 64\}$ms.

For each parameter set, a 2-fold cross-validation was performed by splitting the data into equal-sized halves. The kernels were computed from data in the first half, and the reconstruction error was calculated for the second half. The procedure was then repeated by estimating from the second half and reconstructing the first. The reconstructed spiking vector was computed as $[ \mathbf{\tilde{K}}_{ff} \ \mathbf{\tilde{K}}_{fb}]\mathbf{\tilde{X}}$.

Reconstruction error was defined as:
\[
\text{Error} = \frac{\|\mathbf{U} - [ \mathbf{\tilde{K}}_{ff} \ \mathbf{\tilde{K}}_{fb}]\mathbf{\tilde{X}}\|_2^2}{\|\mathbf{U}\|_2^2}
\]
This estimation procedure was done individually for each trial. The filters plotted in Fig. \ref{fig2} in the main text represent the mean across all trials, with dotted lines indicating mean plus or minus the standard error of the mean.
The number and length of trials for each dataset in Fig. \ref{fig2} in the main text were as follows:
\begin{table}[h]
\centering
\begin{tabular}{ccccc}
\hline
 & Trials & Trial Length & Time Step & Maximum Shift* \\
\hline
2B (Dark) & 135 & 5000 & 2 ms & 100 data points \\
2B (Mid) & 135 & 5000 & 2 ms & 100 data points \\
2B (Light) & 135 & 5000 & 2 ms & 100 data points \\
2C (Low) & 8 & 3000 & 1 ms & 100 data points \\
2C (High) & 8 & 3000 & 1 ms & 100 data points \\
2D (Bar) & 62 & 8270 & 1 ms & 200 data points \\
2D (Fish) & 102 & 17705 & 1 ms & 200 data points \\
2E (Low) & 30 & 5000 & 1 ms & 200 data points \\
2E (High) & 30 & 5000 & 1 ms & 200 data points \\
2F (Each Condition) & 8 & 2375 & 1 ms & 150 data points \\
\hline
\end{tabular}
\caption{* Maximum Shift refers to the maximum number of time steps that the stimulus is shifted when constructing the lagged stimulus matrix.}
\end{table}

\section*{E. Control Analyses}
\subsection*{E.1 Mean Firing Rate}
\begin{figure}[h]
  \centering
\includegraphics[width=.4\textwidth]{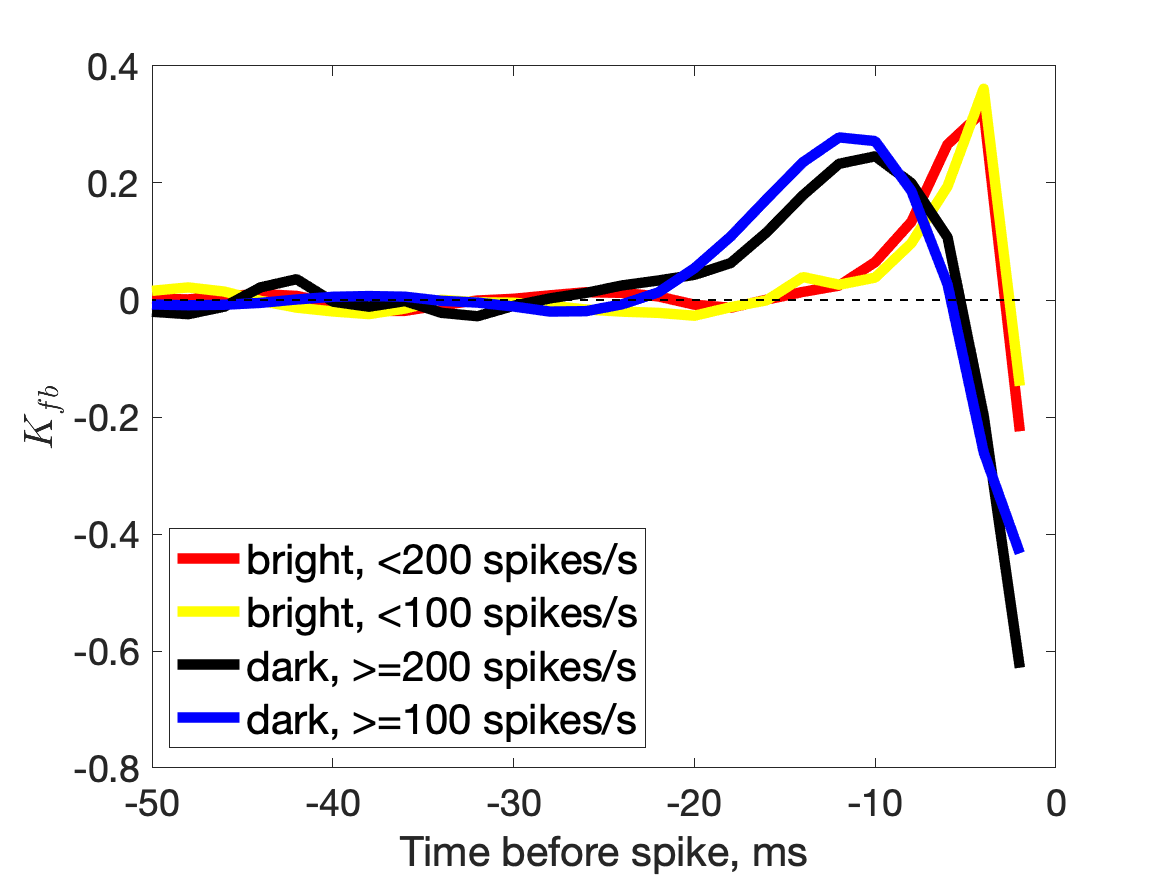}
  \caption{Blowfly H1 feedback filter shape is primarily determined by the statistics of sensory stimuli (background luminance) rather than the H1 firing rate. Even when we analyze data subsets from the bright background experiment with firing rate lower than that in data subsets from the dark background experiment, the difference between the filters is qualitatively similar to Fig. \ref{fig2}\textit{B} in the main text.}
  \label{fig7}
\end{figure}

In the fly H1 dataset spike rates decrease considerably in darker conditions, so we performed supplementary analyses to rule out the possibility that the change in filter shape is simply a consequence of a different firing rate. We repeated the analysis used in Fig. \ref{fig2} in the main text for bright segment of the dataset while keeping only those data where the number of spikes in the lag vector were below a threshold (100 or 200 spikes/s). Similarly, for the dark segment, the analysis was repeated while keeping data points where the number of spikes in the lag vector was above some threshold (100 or 200 spikes/s). As shown in Fig. \ref{fig7}, the filter shape is determined primarily by belonging to the dark/bright segment rather than by the firing rate. Therefore, this control supports our assertion that the filter shapes adapt to the stimulus statistics.
\subsection*{E.2 Sunrise and Sunset}
\begin{figure}[h]
  \centering
\includegraphics[width=.5\textwidth]{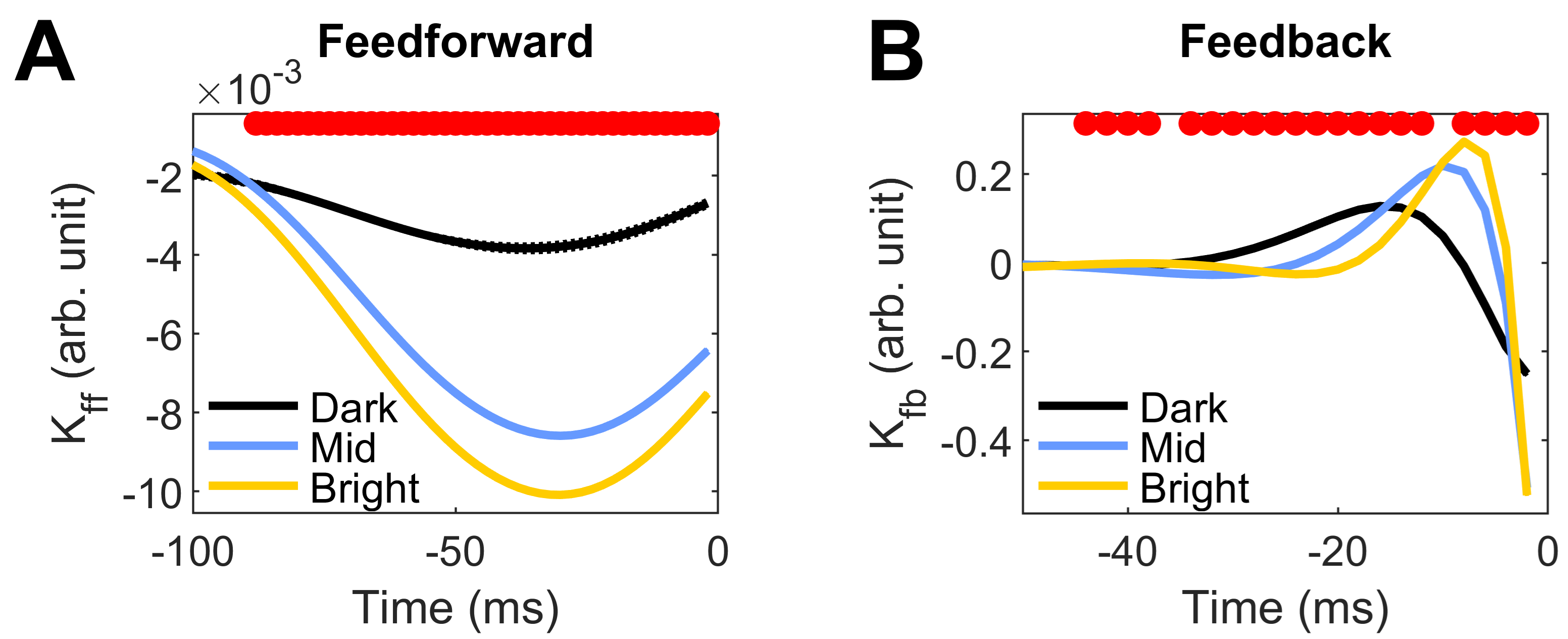}
  \caption{Changes in Blowfly H1 feedforward and feedback filters are due to changes in environmental brightness, not overall recording time. The above filters were obtained from a dataset recorded during sunrise, in which the Dark condition came first, followed by Mid and then Bright. This is in contrast to the analyses in Fig. \ref{fig2}\textit{B} which was recorded during sunset, in which Bright came first, followed by Mid and then Dark. The changes in the filters are qualitatively the same between these conditions, indicating that this is driven by the environmental brightness and not an artifact of overall recording time.}
  \label{fig8}
\end{figure}

A possible explanation for the change in feedforward and feedback filters in the fly H1 dataset is that the the effect was due to the elapsed recording time. To rule this out, we repeated the analysis on a separate H1 recording performed during the time around sunrise rather than sunset. The response profiles were remarkably similar for periods of the experiment with comparable light conditions, Fig. \ref{fig8}, indicating that the change in filter shape is primarily due to an effect of environmental brightness rather than recording time.
\subsection*{E.3 Effect of Z-Scoring Spike Trains}
\begin{figure*}[ht]
  \centering
  \includegraphics[width=\textwidth]{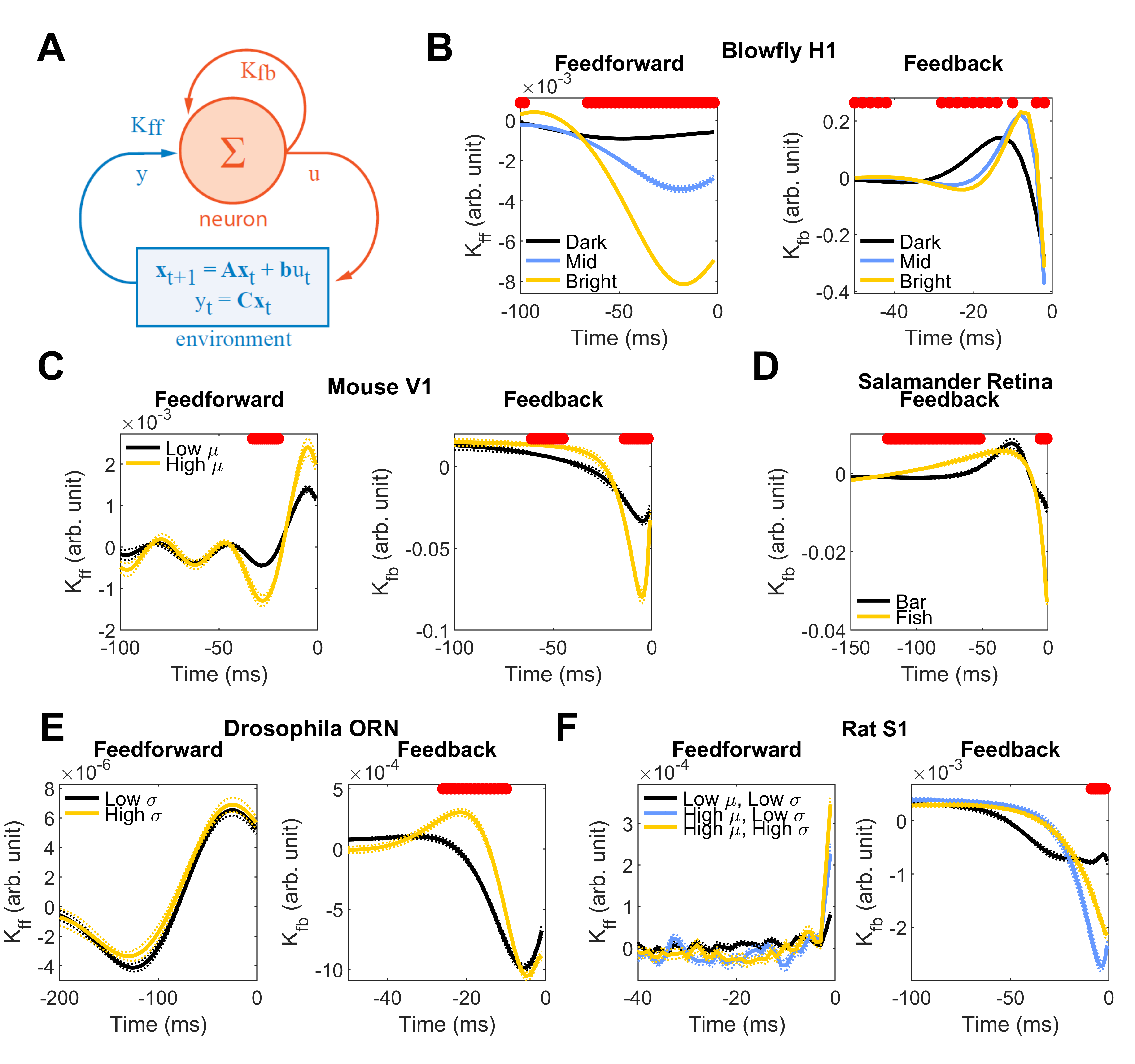}
  \caption{Estimated filters without z-scoring the spiking vector $u(t)$ are qualitatively similar to the ones with z-scoring, Fig. \ref{fig2}, in the main text. Naming and coloring conventions are as in Fig. 3 in the main text. Feedforward filters in \textit{F} were not substantially different from those obtained using the Z-scored data.}
  \label{fig9}
\end{figure*}

Spike trains with different mean firing rates will be affected differently by z-scoring (the first step in Data Preparation). Time bins with spikes will take on larger values at low rates than at high rates. This disparity in the magnitude of the non-zero components of the spike train may affect the shape or scaling of the estimated filters.
To control for this, we re-ran the fitting procedure for all datasets while not z-scoring the spiking vector $\mathbf{U}$, Fig. \ref{fig9}. The filter shapes and differences across conditions are extremely similar to those obtained with z-scoring, aside from a scaling factor for the feedforward filters.

\end{document}